# Negative effective permeability and left-handed materials at optical frequencies

**A. Alù and A. Salandrino**

*University of Pennsylvania, Department of Electrical and Systems Engineering,
200 South 33rd Street, Philadelphia, PA 19104, U.S.A.*

*University of Roma Tre, Department of Applied Electronics,
via della Vasca Navale, 84, Roma, RM 00146, Italy*
*alu@uniroma3.it*

**N. Engheta**

*University of Pennsylvania, Department of Electrical and Systems Engineering,
200 South 33rd Street, Philadelphia, PA 19104, U.S.A.*
*engheta@ee.upenn.edu*

**Abstract:** We present here the design of nano-inclusions made of properly arranged collections of plasmonic metallic nano-particles that may exhibit a resonant magnetic dipole collective response in the visible domain. When such inclusions are embedded in a host medium, they may provide metamaterials with negative effective permeability at optical frequencies. We also show how the same inclusions may provide resonant electric dipole response and, when combining the two effects at the same frequencies, left-handed materials with both negative effective permittivity and permeability may be synthesized in the optical domain with potential applications for imaging and nano-optics applications.

**OCIS codes:** (160.4670) Optical materials; (160.3900) Metals.

## 1. Introduction

The interest in materials with negative effective magnetic, as well as electric, properties has grown considerably in the past several years, mainly due to the recent interest in unconventional characteristics of composite *metamaterials* with both negative permittivity and permeability, also known as left-handed (LH) or double negative (DNG) materials [1]. In the microwave regime, such complex materials have been constructed by embedding arrays of metallic split-ring resonators (SRR) and wires in a host medium and some of their anomalous properties have been experimentally demonstrated [2]. In the near-infrared (IR) and visible regimes, however, synthesizing such LH materials faces certain challenges, since in these frequency regimes the magnetic permeability due to the molecular currents in a material tends to approach to the free space permeability [3], and the straightforward scaling of the metallic SRR down to the optical wavelength indeed encounters related challenges. In particular, in addition to challenges in nanofabrication of ring or loop resonators and small gaps, it should be mentioned that the electric conductivity of metals, on which the resonance of SRR at microwave frequencies depends, behaves differently as the frequency is increased into the IR and visible domains. Following these issues, several novel ideas have been put forward by other researchers to achieve LH materials in the IR and visible regimes. They include the possibility of using coupled plasmonic parallel nano-wires and nano-plates [4]-[8], coupled nano-cones [9], anisotropic waveguides [10], modified SRR in the near-IR region [11]-[12], closely-packed inclusions with negative permittivity and their electrostatic resonances [13], and defects in regular photonic band gap structures [14].

Here we present another approach to design sub-wavelength inclusions that exhibit magnetic dipolar resonant response, and thus provide the possibility of having negative effective magnetic dipole moment, at optical frequencies. The idea is based on the collective resonance of an array of plasmonic nano-particles arranged in a specific pattern (e.g., in a circular pattern) to form a single sub-wavelength "ring inclusion". In this ring, it is not the conventional conduction current (as in the SRR at microwaves) that produces the magnetic dipole moment, but instead it is the plasmonic resonant feature of every nano-particle that induces a circulating "displacement" current around this loop. Unlike the case of the conventional metallic loops or SRRs at the microwave frequencies, here the size of this loop does not directly influence the resonant frequency of the induced magnetic dipole moment, but rather the plasmonic resonant frequency of the nano-particle is the main determining factor for this resonance to happen.

These results may open interesting venues for realizing left-handed metamaterials at optical frequencies, with potential applications in imaging and nano-optics. In the following we justify theoretically our findings and validate the results with numerical simulations and some physical insights.

## 2. Geometry of the nano-ring and its electromagnetic response

Consider $N$ identical nano-spheres with radius $a$ arranged to have their centers located symmetrically on a circle of radius $R$, as depicted in Fig. 1. A Cartesian coordinate system $(x, y, z)$ along with a related spherical coordinate system $(r, \theta, \phi)$ is considered here. We assume $a \ll R \ll \lambda_b$, where $\lambda_b$ is the wavelength in the background medium, i.e., the material where the spheres are embedded. The position vectors of the centers of these nano-particles are described as:

$$\mathbf{r}_n = \hat{\mathbf{x}}\left[R\cos\left((n-1)\frac{2\pi}{N}\right)\right] + \hat{\mathbf{y}}\left[R\sin\left((n-1)\frac{2\pi}{N}\right)\right] = R\,\hat{\mathbf{r}}(\mathbf{r}_n), \tag{1}$$

with $n = 1, 2, .. N$, and $\hat{\mathbf{r}}$ being the spherical radial unit vector.

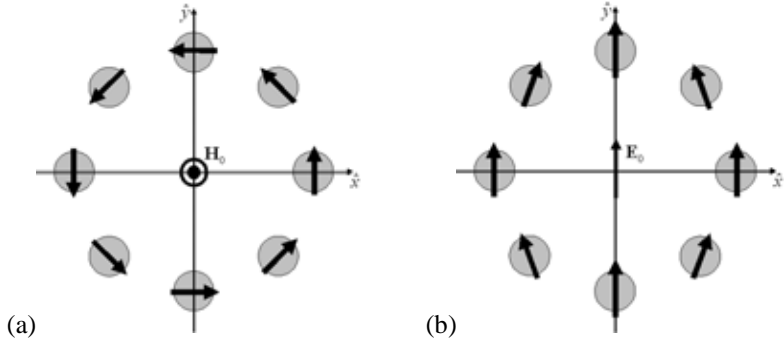

Fig. 1. A circular array of equi-spaced nano-spheres in the x-y plane excited by: (a) a local time-varying magnetic field directed along *z*; (b) a local time-varying electric field directed along *y*. The vectors on each particle indicate the induced electric dipole moments in the two cases.

*2.1 Magnetic dipole moment and effective magnetic permeability*

In order to single out and evaluate the magnetic response of this structure that is needed to evaluate the magnetic polarizability of such a nano-loop, without exciting any noticeable electric response, we excite the system with the following system of $N$ symmetrical plane waves (see e.g., [15]):

$$\mathbf{H}_0 = \hat{\mathbf{z}}\sum_{n=1}^{N}\frac{H_0}{N}e^{i\mathbf{k}_n \cdot \mathbf{r}}, \quad \mathbf{E}_0 = \sum_{n=1}^{N}\mathbf{E}_{0n}e^{i\mathbf{k}_n \cdot \mathbf{r}}, \tag{2}$$

with:

$$\begin{cases} \mathbf{k}_n = -k_b \hat{\mathbf{r}}(\mathbf{r}_n) \\ \mathbf{E}_{0n} = -\dfrac{\eta_b H_0}{N}\hat{\boldsymbol{\varphi}}(\mathbf{r}_n) \end{cases}, \tag{3}$$

where $\eta_b = \sqrt{\mu_0/\varepsilon_b}$ is the background characteristic impedance (i.e., the square root of the ratio between its permeability and its permittivity), $k_b = \omega\sqrt{\varepsilon_b\mu_0}$ is the background wave number, $\hat{\boldsymbol{\varphi}}$ is the spherical azimuth unit vector, and the impinging wave is assumed to be monochromatic with an $e^{-i\omega t}$ time dependence. Moreover, $H_0$ is the magnetic field amplitude at the origin produced by this combined excitation [16].

Since the nano-spheres composing the loop are small compared with the wavelength and the distance between adjacent nano-spheres is taken to be relatively large compared to their diameters, in the first-order approximation the interactions among the particles can be described by their induced electric dipole moments:

$$\mathbf{p}_n = \alpha\,\mathbf{E}_{loc}(\mathbf{r}_n) = p\,\hat{\boldsymbol{\varphi}}(\mathbf{r}_n), \tag{4}$$

where $\alpha$ is the polarizability coefficient of the particles (which is scalar since the nano-spheres are assumed to have an isotropic behavior) and $\mathbf{E}_{loc}(\mathbf{r}_n)$ is the local field at the $n^{th}$ particle location (in the absence of that particle), given by the sum of the incident applied field and the fields scattered by all the other particles at that point. The last equality in (4) holds due to the symmetry of the geometry and of the excitation. In fact, as depicted in Fig. 1(a), the symmetric excitation of Eq. (2) provides an almost spatially uniform time-varying magnetic field around the inclusion, inducing electric dipole moments (due to Faraday's induction) on each nano-particle that "circulate" around the axis of the magnetic field.

The local field $\mathbf{E}_{loc}(\mathbf{r}_n)$ may be written in the following form:

$$\mathbf{E}_{loc}(\mathbf{r}_n) = \mathbf{E}_0(\mathbf{r}_n) + p\sum_{l \neq n}^{N} \underline{\mathbf{Q}}_{ln} \cdot \hat{\boldsymbol{\varphi}}(\mathbf{r}_l), \tag{5}$$

where $\underline{\mathbf{Q}}_{ln}$ is the Green's dyad, as usually defined (see e.g., [17]). Combining (4) and (5) we get:

$$p = \frac{\mathbf{E}_0(\mathbf{r}_n) \cdot \hat{\boldsymbol{\varphi}}(\mathbf{r}_n)}{\alpha^{-1} - \sum_{l \neq n}^{N} \underline{\mathbf{Q}}_{ln} \cdot \hat{\boldsymbol{\varphi}}(\mathbf{r}_l) \cdot \hat{\boldsymbol{\varphi}}(\mathbf{r}_n)}. \tag{6}$$

In the limit of $k_b R \ll 1$, the electromagnetic far-zone fields scattered by such a configuration of nano-particles are given by:

$$\begin{aligned}\mathbf{E} &\simeq -\frac{iNk_b^3 R}{8\pi\varepsilon_b}\frac{e^{ik_b r}}{r} p \sin\theta\,\hat{\boldsymbol{\varphi}} \\ \mathbf{H} &\simeq \frac{iNk_b^3 R}{8\pi\varepsilon_b \eta_b}\frac{e^{ik_b r}}{r} p \sin\theta\,\hat{\boldsymbol{\theta}}\end{aligned}, \tag{7}$$

which by inspection correspond to those radiated by an effective magnetic dipole with amplitude

$$\mathbf{m}_H = \frac{-i\omega p N R}{2}\hat{\mathbf{z}}. \tag{8}$$

It may be verified that the quasi-static multipole expansion of the current distribution $\mathbf{J}_H = \sum_{n=1}^{N} -i\omega\mathbf{p}_n \delta(\mathbf{r}-\mathbf{r}_n)$ yields to a magnetic dipole moment consistent with (8), ensuring, moreover, that electric moments up to the order $N-1$ and even-order magnetic moments up to the order $N$ are identically zero for this excitation. The other non-vanishing higher-order electric and magnetic multipoles are respectively proportional in amplitude to $(k_b R)^{m-1}$ and $(k_b R)^m$, where $m > N$.

The relative strength of the higher-order multipoles provides an insight into how closely packed we may embed many such nano-loops in a host medium in order to form a bulk material with magnetic properties. It is clear that, by increasing $N$ and/or decreasing $(k_b R)$, more higher-order multipoles may be canceled or diminished, and therefore the ratio between the non-vanishing higher-order multipole amplitudes and the magnetic dipole moment may be made sufficiently small to be neglected.

The magnetic polarizability $\alpha_{mm}$ of the nano-ring, which relates the induced magnetic dipole moment to the excitation field through $\mathbf{m}_H = \alpha_{mm} \mathbf{H}_0(\mathbf{0})$, can be calculated from (8) and (6). In the limit of $k_b R \ll 1$, its expression is written in closed form as:

$$\alpha_{mm}^{-1} = \frac{4\varepsilon_b}{Nk_b^2 R^2}\alpha^{-1} - i\left(\frac{k_b^3}{6\pi} - \frac{2k_b}{3\pi NR^2}\right) + \frac{1}{16\pi Nk_b^2 R^5}\sum_{l \neq n}^{N} \frac{3 + \cos[2\pi(l-n)/N]}{\left|\sin[\pi(l-n)/N]\right|^3}. \tag{9}$$

For a small homogeneous isolated single nano-sphere of permittivity $\varepsilon$ the expression for $\alpha$ may be given as [18]:

$$\alpha = \left[\left(4\pi\varepsilon_b a^3 \frac{\varepsilon - \varepsilon_b}{\varepsilon + 2\varepsilon_b}\right)^{-1} - i\frac{k_b^3}{6\pi\varepsilon_b}\right]^{-1}. \tag{10}$$

As evident from (9), the magnetic polarizability may hit the resonance for values close to the resonance of the single nano-sphere electric polarizability, which happens at frequencies for which $\varepsilon \simeq -2\varepsilon_b$. In particular, the resonance in (9) is slightly shifted along the frequency axis with respect to (10), due to the coupling terms represented by the summation.

As long as the loop inclusion is small compared to the wavelength ($k_b R \ll 1$) and therefore (9) is valid and describes sufficiently well the electromagnetic properties of the "loop" inclusion, it is possible to embed many such loops in a host medium in order to synthesize a composite material with resonant magnetic properties at optical frequencies. The effective permeability of such a composite can be obtained using the effective medium theory for a case of randomly located inclusions and for a case of 3-D periodic arrangement of such inclusions [19], and is expressed as:

$$\mu_{eff}^{(r)} = \mu_0\left(1 + \frac{1}{N_d^{-1}\alpha_{mm}^{-1} - 1/3}\right), \quad \mu_{eff}^{(p)} = \mu_0\left(1 + \frac{1}{N_d^{-1}\left[\alpha_{mm}^{-1} + i(k_b^3/6\pi)\right] - 1/3}\right), \tag{11}$$

for the random distribution and the regular 3-D periodic distribution of loops, respectively. Here $N_d$ is the number density of loop inclusions per unit volume.

Fig. 2 shows, as examples, the behavior of $\mu_{eff}^{(p)}$ for two periodic lattices designed to exhibit a negative permeability in the optical domain. The loops have been designed using silver nano-spheres with the appropriate Drude model for their permittivity in this frequency range [20]-[21]. Realistic material loss has been included in this model. The loop has been embedded in a glass substrate with $\varepsilon_b = 2.2\varepsilon_0$ (note that the background material represents an additional degree of freedom that may move the resonant frequency to the specific desired value). It is worth noting how the magnetic resonance is appreciable over a relatively wide range of frequencies. For a higher number of spheres per loop the resonance is generally stronger, as clearly evident by comparing Fig. 2(a) and Fig. 2(b), even though the resonance may shift in frequency, due to the different coupling among the nano-particles.

We have verified and validated our analytical results with full-wave numerical solutions using a finite-integration technique commercial simulator (CST Microwave Studio™ [22]). Here we include a movie (Fig. 3) of the time domain electric field distribution at 655 THz for the particle of Fig. 2(b), confirming the resonant behavior of this nano-ring of four particles at its magnetic resonant frequency. You notice how the displacement current circulates around the loop and closes itself in a ring, analogously to what happens in a SRR at microwave frequencies for the circulating conduction current. At this specific frequency (very close to the one predicted by our analytical model) this "magnetic resonance" dominates any other response of the collection of nano-particles, even though all the employed materials are inherently non-magnetic.

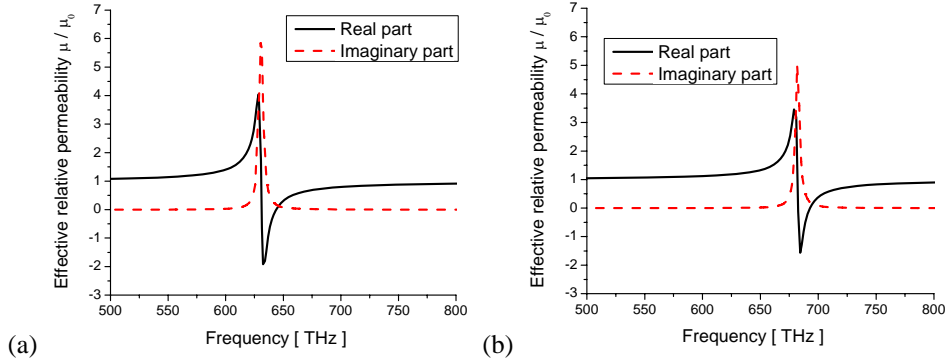

Fig. 2. Effective relative magnetic permeability $\mu_{eff}^{(p)}/\mu_0$ for bulk media with the following parameters for the geometry of nano-inclusion: (a) $R=40\,nm$, $a=16\,nm$, $N=6$, $N_d=(108\,nm)^{-3}$; (b) $R=38\,nm$, $a=16\,nm$, $N=4$, $N_d=(95\,nm)^{-3}$. Following [20], in this range of frequency the permittivity of silver has been assumed to follow the Drude model with $f_p^{Ag}=2175\,THz$, $f_\tau^{Ag}=4.35\,THz$ and $\varepsilon_\infty=5\varepsilon_0$. The background material in this example is glass with $\varepsilon_b=2.2\varepsilon_0$.

Similar results may be obtained by removing the hypothesis of isotropic particles that form the loop. For symmetry reasons, in fact, following the previous analysis, we may employ particles with "anisotropic" polarizabilities (e.g., ellipsoids, nano-rods, nano-discs), provided that their relative orientation is parallel to $\hat{\boldsymbol{\varphi}}(\mathbf{r}_n)$. This may provide further degrees of freedom to allow designing a bulk material with magnetic response having the resonance at a desired frequency.

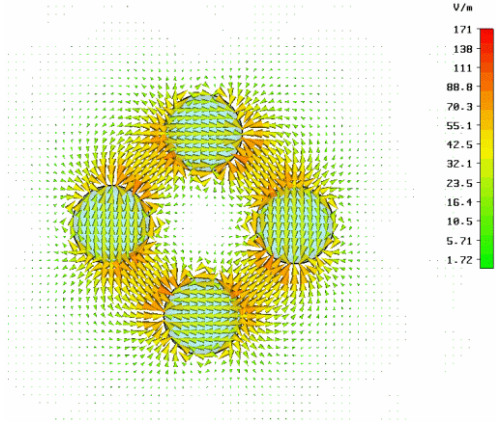

Fig. 3. (2.5 MB) Movie (simulated with CST Microwave Studio[TM] [22]) of the time-domain electric field distribution for the nano-ring of Fig. 2(b), composed of four plasmonic nano-spheres, at $f = 655 THz$.

To summarize, here we have shown how the nano-ring of Fig. 1 may indeed exhibit a magnetic resonant behavior at optical frequencies, even though each of the particles composing the loop is non-magnetic. This is somehow consistent with, and analogous to, the other different geometries in which coupled plasmonic nano-particles have been properly designed and adjusted to induce magnetic plasmonic antisymmetric resonances (see e.g., [4]-[9]). We may speculate that the particular geometry studied here may induce a relatively stronger magnetic effect as compared with the other (higher order) multipoles of this geometry, as one can see from the analysis here. This is due to the circulating displacement current that closes itself in a *circular* loop, guided by the specific geometry of the plasmonic nano-particles, representing effectively a magnetic dipole at optical frequencies (see Fig. 3). It is important to reiterate, moreover, that in this geometry it is not mainly the size of the ring, i.e., its diameter, that causes a resonance in this magnetic response, but rather its resonance depends in most part on the plasmonic resonance of the constituent particles, as Eq. (9) shows. This allows an effective design of sub-wavelength resonant magnetic nano-loops at optical frequencies.

*2.2 Electric dipole moment and effective electric permittivity*

The electric response of the nano-inclusion of Fig. 1 may be calculated in an analogous way [19]. This time the composed excitation should be:

$$\begin{aligned} \mathbf{E}_0 &= E_0 \cos(k_b x)\hat{\mathbf{y}} \\ \mathbf{H}_0 &= \frac{iE_0 \sin(k_b x)}{\eta_b}\hat{\mathbf{z}} \end{aligned}, \qquad (12)$$

which corresponds to having two symmetric, counter-propagating plane waves added together in phase at the origin, in order to isolate the electric response of the inclusion without having any noticeable magnetic dipole. For small enough loops, the induced total electric dipole moment will be proportional to the electric polarizability $\alpha_{ee}^y$ of the loop. As will be seen below, due to the lack of symmetry, the loop exhibits an anisotropic response for its electric polarizability even in the x-y plane. By increasing the number $N$ of nano-particles composing the loop, however, this planar anisotropy diminishes.

The dipole moment induced on the $n-$th sphere may be evaluated through the vectorial relation:

$$\mathbf{p}_n = \alpha \mathbf{E}_{loc} = \alpha \left[ \mathbf{E}_0(\mathbf{r}_n) + \sum_{l \neq n}^{N} \underline{\mathbf{Q}}_{ln} \cdot \mathbf{p}_l \right]. \tag{13}$$

A system of $N$ equations (13), for $n = 1..N$, may be solved numerically to derive the induced dipole moments $\mathbf{p}_n$. In the limit of $k_b R \ll 1$, the multipole expansion of such a distribution is dominated by the effective dipole moment $\mathbf{p}_E^{(1)} = \sum_{n=1}^{N} \mathbf{p}_n$, which, due to the symmetry, is parallel to the applied field $\mathbf{E}_0$. The induced dipole distribution for this case is sketched in Fig. 1(b). The related polarizability factor $\alpha_{ee}^{y}$, which satisfies the relation $\mathbf{p}_E = \alpha_{ee}^{y} \mathbf{E}_0(\mathbf{0})$, may be straightforwardly calculated numerically and analogous results may be obtained for the quantity $\alpha_{ee}^{x}$, for an electric field excitation polarized along $\hat{\mathbf{x}}$. The two quantities are expected to be the same for $N$ being a multiple of four, and increasingly more similar for higher values of $N$.

The effective permittivity for the bulk medium is given by the following expressions, analogous to (11):

$$\varepsilon_{eff}^{(r)} = \varepsilon_0 \left( 1 + \frac{1}{\varepsilon_0 N_d^{-1} \alpha_{ee}^{-1} - 1/3} \right), \quad \varepsilon_{eff}^{(p)} = \varepsilon_0 \left( 1 + \frac{1}{\varepsilon_0 N_d^{-1} \left[ \alpha_{ee}^{-1} + i \left( k_0^3 / 6\pi\varepsilon_0 \right) \right] - 1/3} \right). \tag{14}$$

Fig. 4 reports the effective permittivity calculated for the same composite media simulated in Fig. 2. It is interesting to note how for a larger number of spheres per loop it is possible to achieve multiple resonances, as in Fig. 4(a), one of which is mainly due to the local plasmonic resonance of each nano-sphere and the others are due to the coupling among the resonant nano-spheres. The dotted lines in Fig. 4 represent the effective permittivity of another composite medium that could be formed by embedding the same number density of nano-spheres when these spheres are dispersed in a regular periodic lattice, not collected in loops as in our case. As can be seen, as the number of spheres per loop increases, the two curves become more different, since this loop effect becomes more pronounced. It is interesting to point out however that the permittivity resonance described in Fig. 4 does not differ conceptually from the permittivity behavior of randomly displaced particles, whereas the magnetic resonance presented in Fig. 2 mainly derives from the loop-shaped inclusion here presented.

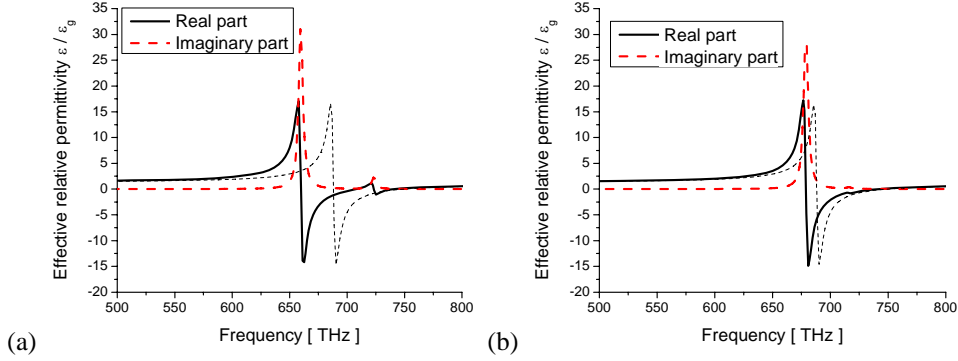

(a) (b)

Fig. 4. Effective relative electric permittivity $\varepsilon_{eff}^{(p)}/\varepsilon_0$ for the bulk media with the parameters of Fig. 2(a) and 2(b). The dot-lines represent the effective permittivity of another bulk medium that could be formed by embedding the same number density of silver nano-spheres as in the simulations, but in a regular periodic cubic lattice, not collected in loop arrangement.

To conclude this section, it is worthy to reiterate, following [16], that the electric and magnetic responses of the loop inclusion shown in Fig. 1 have been determined separately by impressing two distinct excitations, composed of collections of plane waves that induce separately an almost uniform electric or magnetic field on the loop, respectively. This technique, commonly utilized (see e.g., [15]) to determine independently the effective electric and magnetic polarizabilities of an isolated particle, determines its responses to any kind of excitation. In particular, when a single plane wave impinges on this sub-wavelength particle, its total scattering response is given by the contributions of the electric and magnetic fields locally exciting the inclusion and thus inducing, under the assumption of sub-wavelength size of the particle, the equivalent electric and magnetic dipoles evaluated in our analysis. From these quantities, the effective permittivity, permeability and index of refraction of a metamaterial formed by a collection of these inclusions can be directly determined using the standard techniques.

## 3. Left-handed metamaterials at optical frequencies

It is interesting to point out how the metamaterial synthesized in Fig. 2(b) and Fig. 4(b) shows a range of frequency in the visible in which both effective permittivity and permeability have negative real parts simultaneously. This happens since the small number of spheres per loop (four in the example) does not sensibly shift the resonance position along the frequency axis for the magnetic permeability. Therefore the two resonances might happen around the plasmonic resonance of the single nano-spheres (around $\varepsilon = -2\varepsilon_0$). This overlapping of the two resonances provides us with the possibility to synthesize an effective double-negative (or LH) material in this frequency regime. The results are reported in Fig. 5, where the effective index of refraction is plotted as a function of frequency. We notice how there is a range of frequency in which this metamaterial may have negative refraction with reasonably low losses.

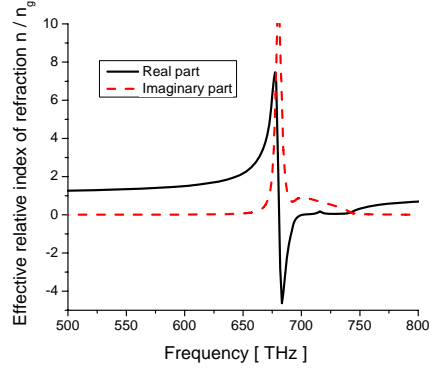

Fig. 5. Effective index of refraction for the material composed as in Fig. 2(b) and Fig. 4(b)

As a final example, we have designed a similar layout to work with a different background material, i.e., SiC with permittivity $\varepsilon_b = 6.5\varepsilon_0$. The results for the expected effective permeability, permittivity and index of refraction are reported in Fig. 6. The plots show how by adjusting the background permittivity the resonance may be tuned at the desired frequency for getting a negative index of refraction with reasonable losses.

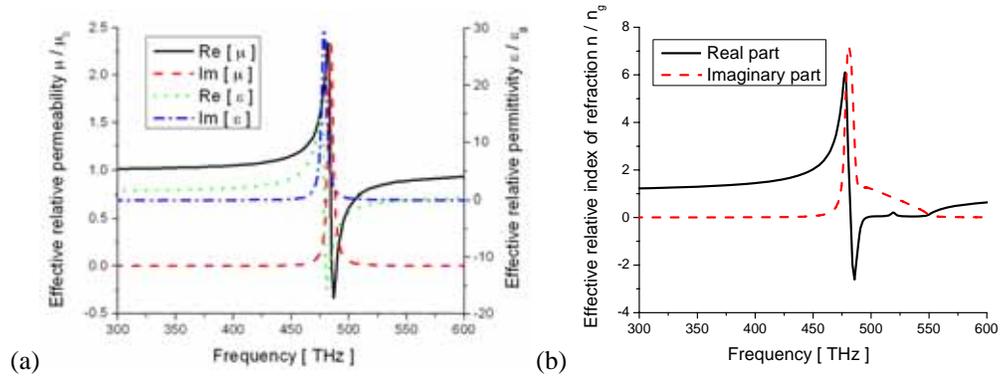

(a) (b)

Fig. 6. (a) Effective magnetic permeability $\mu_{eff}^{(p)}/\mu_0$, and (b) effective electric permittivity $\varepsilon_{eff}^{(p)}/\varepsilon_0$, for the following parameters: $R = 22\,nm$, $a = 9.5\,nm$, $N = 4$, $N_d = (55\,nm)^{-3}$. The background material here is SiC with $\varepsilon_b = 6.5\varepsilon_0$.

## 4. Conclusion

We have shown how it is indeed possible to realize a resonant magnetic response in properly designed nano-inclusions at optical frequency. Relying on the plasmonic resonances of individual metallic nano-particles and arranging the geometry of composite inclusions to resemble magnetic nano-loops, negative-permeability and left-handed metamaterials with a negative index of refraction may be synthesized in the visible domain, with interesting applications for imaging and nano-optics applications.

## Acknowledgments

This work is supported in part by the U.S. Air Force Office of Scientific Research (AFOSR) grant number FA9550-05-1-0442. Andrea Alù has been partially supported by the